\documentclass[12pt,english]{extarticle}

\usepackage{amsmath}

\usepackage{amsfonts}

\usepackage{amssymb}
\usepackage{color,soul}
\usepackage{graphicx}
\usepackage{empheq}
\usepackage{verbatim}

\usepackage{subfigure}

\usepackage{lipsum}

\usepackage[hscale=0.8,vscale=0.9]{geometry}

\usepackage[numbers,sort&compress]{natbib}

\usepackage{eso-pic}

\newcommand\BackgroundPic{
    \put(0,0){
    \parbox[b][\paperheight]{\paperwidth}{%
    \vfill
    \centering
    \includegraphics[width=\paperwidth,height=\paperheight]{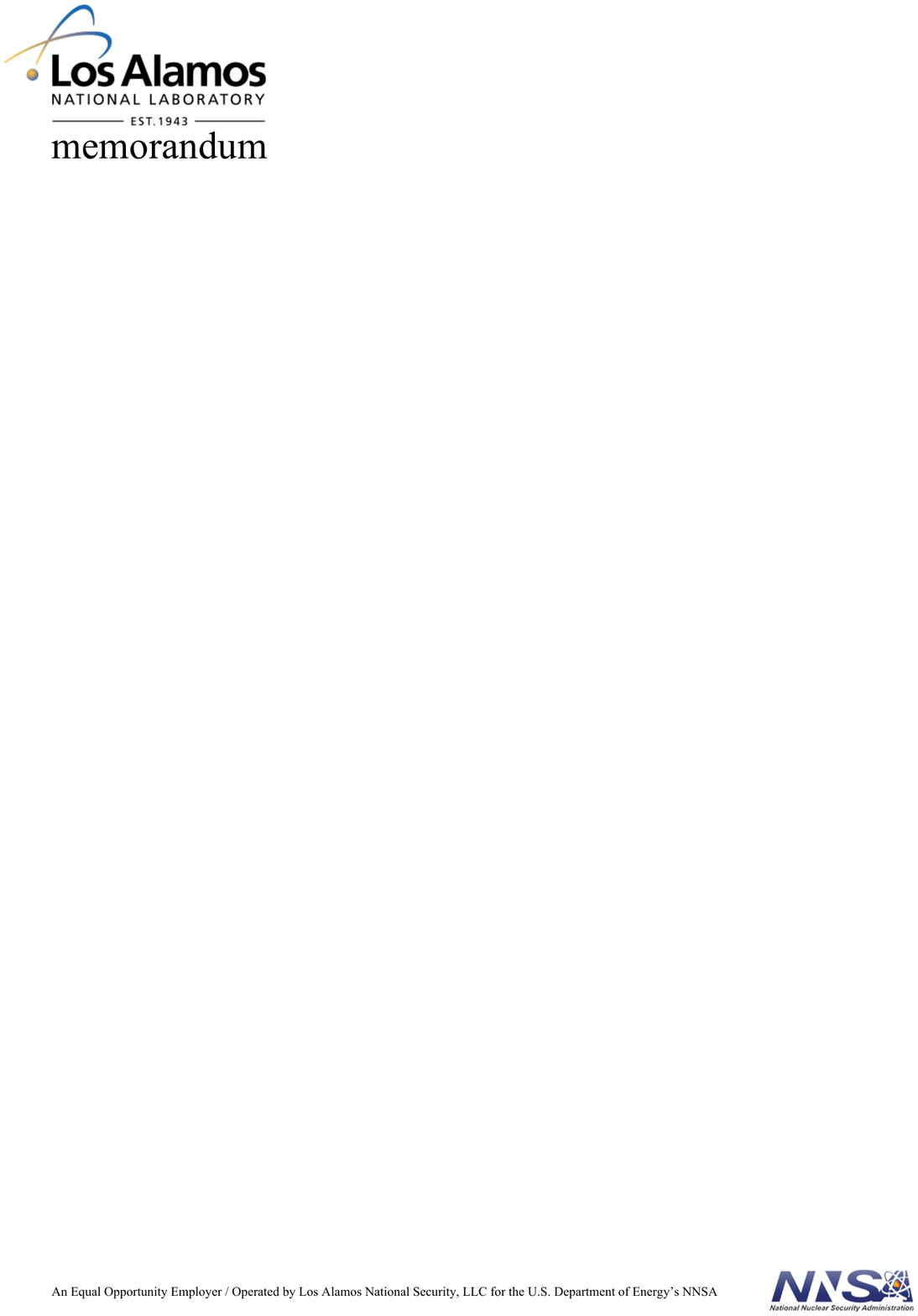}
    \vfill
    }}}

\makeatletter
\renewcommand{\maketitle}{\bgroup\setlength{\parindent}{0pt}
\begin{flushleft}
  \textbf{\@title}

  \@author
\end{flushleft}\egroup
}
\makeatother

\begin{document}

\AddToShipoutPicture*{\BackgroundPic}

\vspace*{3\baselineskip}\vspace{-\parskip}

\begin{flushright}
\text{\textit{To:}~~ICF program and TBI project}~~~~\\
\textit{From:}~~Grigory Kagan (kagan@lanl.gov)\\
\textit{Date:}~~December 10, 2016~~~~~~~~~~~~~~~~~~~
\end{flushright}

\vspace*{2\baselineskip}\vspace{-\parskip}

\title{\large\bf Subject: Comparison of transport coefficients for weakly coupled multi-component plasmas obtained with different formalisms}

{\let\newpage\relax\maketitle}

\date{}


\section{\large\bf Overview}

In recent years LANL has been supporting a substantial body of research on ion multi-species effects. The cornerstone of both qualitative and quantitative consideration of the relevant issues is the transport properties of plasmas with multiple ion species. A number of well established formalisms for weakly coupled multi-component plasmas with arbitrary number of ion species can be found in literature and used readily to obtain all the transport coefficients of interest~\cite{hirschfelder1954molecular, devoto1966transport, ferziger1972mathematical, zhdanov2002transport}. In particular, Kagan \& Tang and Kagan, Baalrud \& Daligault~\cite{kagan2014thermo, kagan2015TBI, kagan2016influence} used existing formalisms by Zhdanov~\cite{zhdanov2002transport} and Ferziger \& Kaper~\cite{ferziger1972mathematical}. On the other hand, in their subsequent work Simakov \& Molvig~\cite{simakov2016hydrodynamic-1,simakov2016hydrodynamic-2} developed a new, self-consistent formalism based on the properly ordered perturbation theory.

It can be noticed, however, that all the existing and the newly developed formalisms for weakly coupled plasmas rely on the same physical assumptions and mathematical approximations: the particles are assumed to interact via Debye shielded Coulomb potential and the linearized kinetic equation is solved by expanding the correction to the species' distribution functions over a set of orthogonal polynomials. In different sources calculation starts with either Boltzmann~\cite{hirschfelder1954molecular, devoto1966transport, ferziger1972mathematical} or Fokker-Planck~\cite{zhdanov2002transport,simakov2016hydrodynamic-1} kinetic equations, but the above mentioned assumption of the Debye shielded Coulomb potential  makes the two equations equivalent. These calculations  utilize either the so-called ``Chapman-Enskog"\cite{hirschfelder1954molecular, devoto1966transport, ferziger1972mathematical,simakov2016hydrodynamic-1}  or ``Grad"~\cite{zhdanov2002transport} methods to solve for the distribution functions, but it is straightforward to observe that  the same, Sonine orthogonal polynomials are employed in both types of the calculations. Hence,  local transport coefficients obtained with all these formalisms must be identical.

The direct comparison between diffusion coefficients obtained with Zhdanov and Ferziger \& Kaper formalisms was demonstrated in Ref.~\cite{kagan2015TBI} to find them identical indeed. Comparison between the diffusion coefficients obtained with Zhdanov and Simakov \& Molvig's formalisms was demonstrated in Ref.~\cite{simakov2006verification} to also find them identical. Since these results have not been distributed to public, in this Note we reproduce the comparison for the diffusion coefficients and demonstrate that results for the electron and remaining ion transport coefficients for weakly coupled plasmas are identical as well.

\section{\large\bf Electron transport}

Electron transport coefficients are provided in Section 8.2 of Ref.~\cite{zhdanov2002transport} and their counterparts obtained with Simakov \& Molvig's formalism were presented in Ref.~\cite{simakov2014electron}. Both sources make the observation that the electron transport coefficients depend on the effective ion charge $Z_{\bf eff}$ only and provide explicit expressions in terms of $Z_{\bf eff}$. Zhdanov addresses a more general case of magnetized plasmas, so to retrieve the unmagnetized  results one should take the longitudinal transport coefficients from~\cite{zhdanov2002transport}. With this notion the electron-ion dynamic friction and thermal force coefficients and the electron heat conductivity by Zhdanov are written as
\begin{align}
\label{eq: heat-forces}
\alpha_{||} &= 1 - \frac{0.22 + 0.73 Z^{-1}}{0.31+  1.20 Z^{-1} + 0.41 Z^{-2}},\\
\beta_{||} &= \frac{0.47 + 0.94 Z^{-1}}{0.31+  1.20 Z^{-1} + 0.41 Z^{-2}},\\
\gamma_{||} &= \frac{3.9 + 2.3 Z^{-1}}{0.31+ 1.20 Z^{-1} + 0.41 Z^{-2}},
\end{align}
respectively.

In the same Chapman-Enskog approximation Simakov \& Molvig find
\begin{align}
\label{eq: heat-forces}
\alpha_{0} &= \frac{4 (16 Z^2 + 61 \sqrt{2} Z +72 ) }{217 Z^2 + 604 \sqrt{2} Z +288 },\\
\beta_{0} &= \frac{30 Z (11Z+15\sqrt{2}) }{217 Z^2 + 604 \sqrt{2} Z +288},\\
\gamma_{0} &= \frac{25 Z (433 Z+180 \sqrt{2}) }{4  (217 Z^2 + 604 \sqrt{2) } Z +288},
\end{align}
respectively.

Zhdanov finds the dimensionless electron viscosity coefficient to be
\begin{align}
\label{eq: heat-forces}
\eta_e^{({0})} = \frac{1.46 + 1.04 Z^{-1}}{0.82+ + 1.82 Z^{-1} + 0.72 Z^{-2}}
\end{align}
and Simakov \& Molvig find it to be
\begin{align}
\label{eq: heat-forces}
\epsilon_0 = \frac{5 Z (408 Z + 205 \sqrt{2} ) }{6 (192 Z^2 + 301 \sqrt{2} Z +178) }.
\end{align}
It is straightforward to observe that for any given coefficient Simakov \& Molvig's expression is identical to its Zhdanov's counterpart except for Zhdanov evaluates the square roots in decimals.

\section{\large\bf Ion heat conductivity and viscosity}

The ion transport coefficients are usually presented in the implicit form---by providing a set of linear algebraic equations, whose solution gives the coefficient(s) of interest. In particular, this is how the ion heat conductivities and viscosities are given by the Simakov \& Molvig, Zhdanov and Ferziger \& Kaper formalisms. The first two formalisms consider the case of a weakly coupled plasma specifically, whereas the Ferziger \& Kaper formalism gives the more general results for an arbitrary binary interaction potential. The information about the interaction potential enters transport coefficients through the standard gas-kinetic cross-sections, the so-called ``$\Omega$-integrals". To apply this formalism to a weakly coupled ionic mixture one needs to insert into the appropriate expressions the $\Omega$-integrals for the Debye shielded Coulomb potential, which are given by 
\begin{align}
\label{eq: Omega-integrals}
\Omega_{i,j}^{(l,r)} = l(r-1)! \frac{  \pi^{1/2} Z_i^2 Z_j^2 \ln \Lambda  }{\mu_{ij}^{1/2} (2T)^{3/2}},
\end{align}
where subscripts $i$ and $j$ denote the ion species and superscript $(l,r)$ the order of the $\Omega$-integral.
Otherwise, formalisms by Simakov \& Molvig and Ferziger \& Kaper are structurally identical and the most natural to compare.

\begin{figure}[h!]
\includegraphics[scale=0.7]{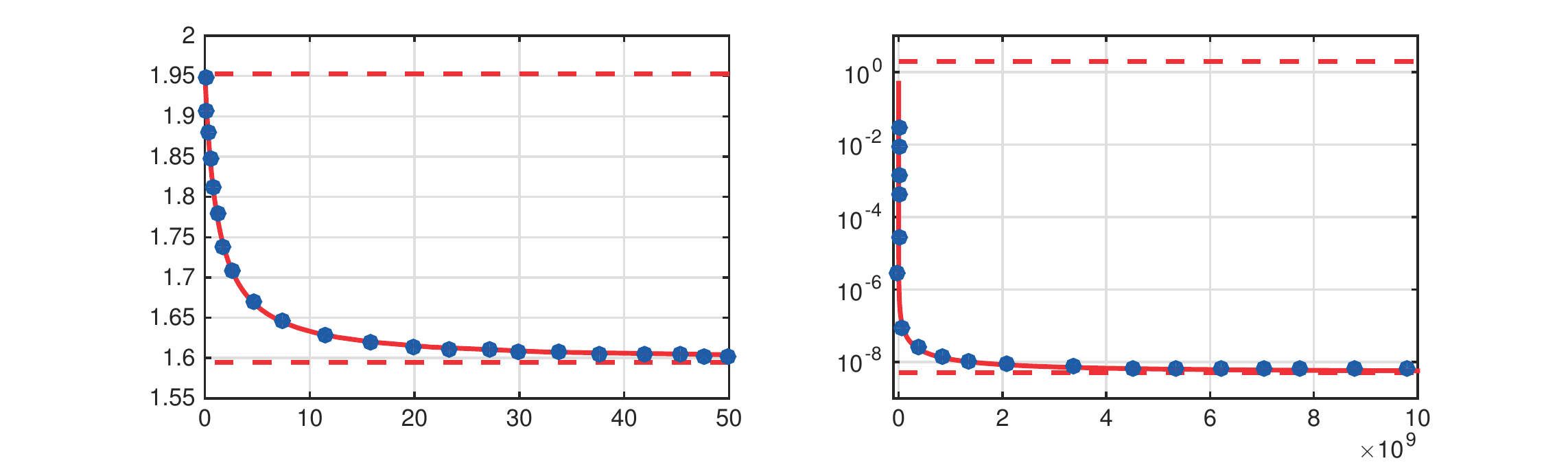}
\caption{The normalized ion heat conductivities for the DT (left) and DAu (right) mixtures as functions of the relative density of the ion species. The definitions of the relative density for the DT and DAu cases are taken to be the same as in Figs.~2 and 5 of Ref.~\cite{simakov2016hydrodynamic-2}, respectively, to facilitate the comparison. Solid red lines show the results obtained with the Ferziger \& Kaper formalism and blue circles show the data obtained by digitizing  Figs.~2 and 5 of  Ref.~\cite{simakov2016hydrodynamic-2}.}
\label{fig: heat} 
\end{figure}

In particular, the ion heat conductivity is calculated as follows. First, the coefficients $\Lambda_{ij}^{pq}$ for the linear set of algebraic equations (6.4-32) are obtained from Eq.~(6.4-15), where the expressions for the bracket integrals in terms of $\Omega_{i,j}^{(l,r)}$ are given in  Table 7.5. With the help of Eq.~(\ref{eq: Omega-integrals}) of this Note, this gives $\Lambda_{ij}^{pq}$ in terms of the ion species densities $n_i$, particle masses $m_i$ and charge numbers $Z_i$. The set of equations~(6.4-32) is then solved for the matrix of coefficients $a_{j,q}^{(n)}$ and the heat conductivity $\lambda'$ is recovered from Eq.~(6.4-45). To compare the results with their counterparts obtained with Simakov \& Molvig formalism we digitize the data from Figs. 2 and 5 of Ref.~\cite{simakov2016hydrodynamic-2} showing the dimensionless heat conductivities for the DT and DAu mixtures, respectively, and normalize $\lambda'$ by Ferziger \& Kaper according to  Eq.~(6) of Ref.~\cite{simakov2016hydrodynamic-2}. The two results are shown in Fig.~\ref{fig: heat} of this Note demonstrating that the two formalisms give identical predictions for the heat conductivity as expected.

\begin{figure}[h!]
\includegraphics[scale=0.7]{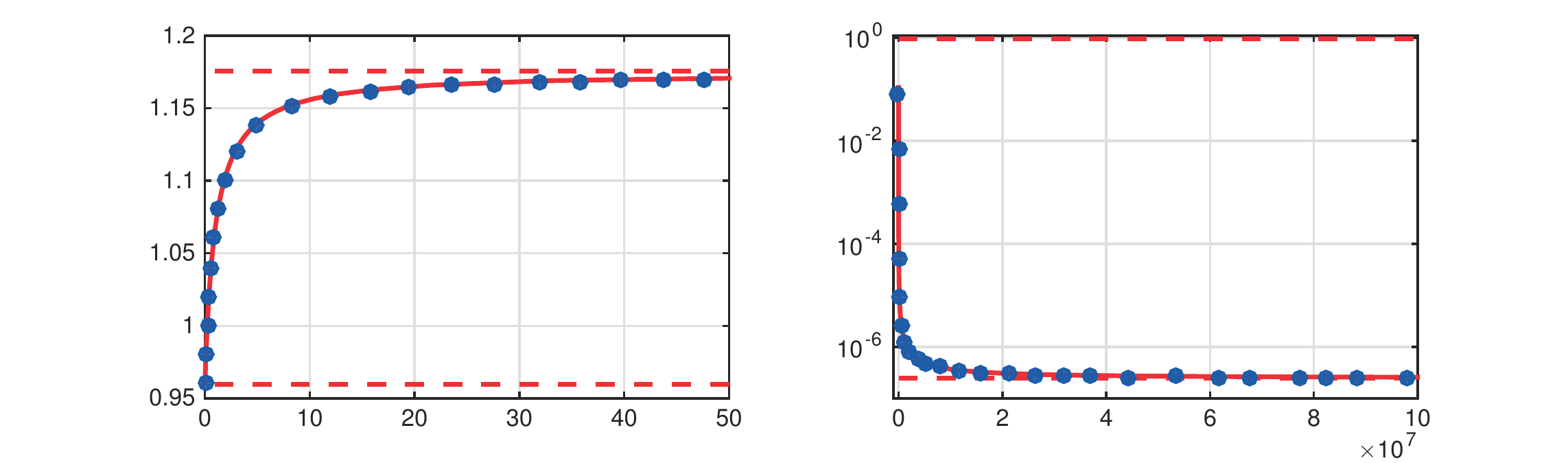}
\caption{The normalized ion viscosities for the DT (left) and DAu (right) mixtures as functions of the relative density of the ion species. The definitions of the relative density for the DT and DAu cases are taken to be the same as in Figs.~3 and 6 of Ref.~\cite{simakov2016hydrodynamic-2}, respectively, to facilitate the comparison. Solid red lines show the results obtained with the Ferziger \& Kaper formalism and blue circles show the data obtained by digitizing  Figs.~3 and 6 of  Ref.~\cite{simakov2016hydrodynamic-2}.}
\label{fig: visc} 
\end{figure}

The viscosity is calculated as follows. First, the coefficients $H_{ij}^{pq}$ for the linear set of algebraic equations (6.4-39) are obtained from Eq.~(6.4-36), where the expressions for the bracket integrals in terms of $\Omega_{i,j}^{(l,r)}$ are given in  Table 7.6. With the help of Eq.~(\ref{eq: Omega-integrals}) of this Note, this gives $H_{ij}^{pq}$ in terms of the ion species densities $n_i$, particle masses $m_i$ and charge numbers $Z_i$. The set of equations~(6.4-39) is then solved for the matrix of coefficients $b_{j,q}^{(n)}$ and the viscosity $\eta$ is recovered from Eq.~(6.4-47). To compare the results with their counterparts obtained with Simakov \& Molvig formalism we digitize the data from Figs. 3 and 6 of Ref.~\cite{simakov2016hydrodynamic-2} showing the dimensionless viscosities for the DT and DAu mixtures, respectively, and normalize $\eta$ by Ferziger \& Kaper according to  Eq.~(7) of Ref.~\cite{simakov2016hydrodynamic-2}. The two results are shown in Fig.~\ref{fig: visc} of this Note demonstrating that the two formalisms give identical predictions for the viscosity as well.

\section{\large\bf Ion diffusion}

Finally, for this Note to be self-contained here we reproduce the comparison for the DT mixture, which was considered in both Ref.~\cite{kagan2014thermo} by Kagan \& Tang and in the subsequent work by Simakov \& Molvig~\cite{simakov2016hydrodynamic-2}. In Ref.~\cite{kagan2014thermo}, the diffusive mass flux is written in the form
\begin{equation}
\label{eq: canonical-flux}
\vec{i} =
- \rho D \Bigl( \nabla c +k_p \nabla \log{p_i} + \frac{e k_E}{T_i}\nabla \Phi + k_T^{(i)} \nabla \log{T_i}  + k_T^{(e)} \nabla \log{T_e}\Bigr),
\end{equation} 
where
\begin{align}
\label{eq: diffusion-coeff}
&D= \frac{\rho T_i}{A_{lh}\mu_{lh} n_l \nu_{lh}} \times \frac{c(1-c)}{cm_h+(1-c)m_l}, \\
\label{eq: baro-diff-ratio}
&k_p = c(1-c)(m_h-m_l)\Bigl( \frac{c}{m_l} +  \frac{1-c}{m_h}    \Bigr),\\
\label{eq: electro-diff-ratio}
&k_E = m_lm_h c(1-c)  \Bigl( \frac{c}{m_l} +  \frac{1-c}{m_h}    \Bigr)  \Bigl( \frac{Z_l}{m_l} - \frac{Z_h}{m_h}  \Bigr),\\
\label{eq: thermo-diff-ratio-el}
&k_T^{(e)} = - m_lm_h c(1-c) \Bigl( \frac{c}{m_l} +  \frac{1-c}{m_h}    \Bigr) \Bigl( \frac{Z_l^2}{m_l}  - \frac{Z_h^2}{m_h}  \Bigr)
\frac{T_e  }{T_i } \frac{\beta_{||}}{Z_{\bf eff}}.
\end{align}
The thermo-diffusion ratio  $k_T^{(i)}$ was evaluated numerically as well as the dynamic friction coefficient $A_{lh}$ needed to retrieve the classical diffusion coefficient $D$ from Eq.~(\ref{eq: diffusion-coeff}) and presented in Figs. 2 and 1, respectively of Ref.~\cite{kagan2014thermo}. In the above equations $c$ is the light species mass fraction, $\Phi$ is the electrostatic potential and $\nu_{lh}$ is the collision frequency between the ion species. 
\begin{figure}[h!]
\includegraphics[scale=0.7]{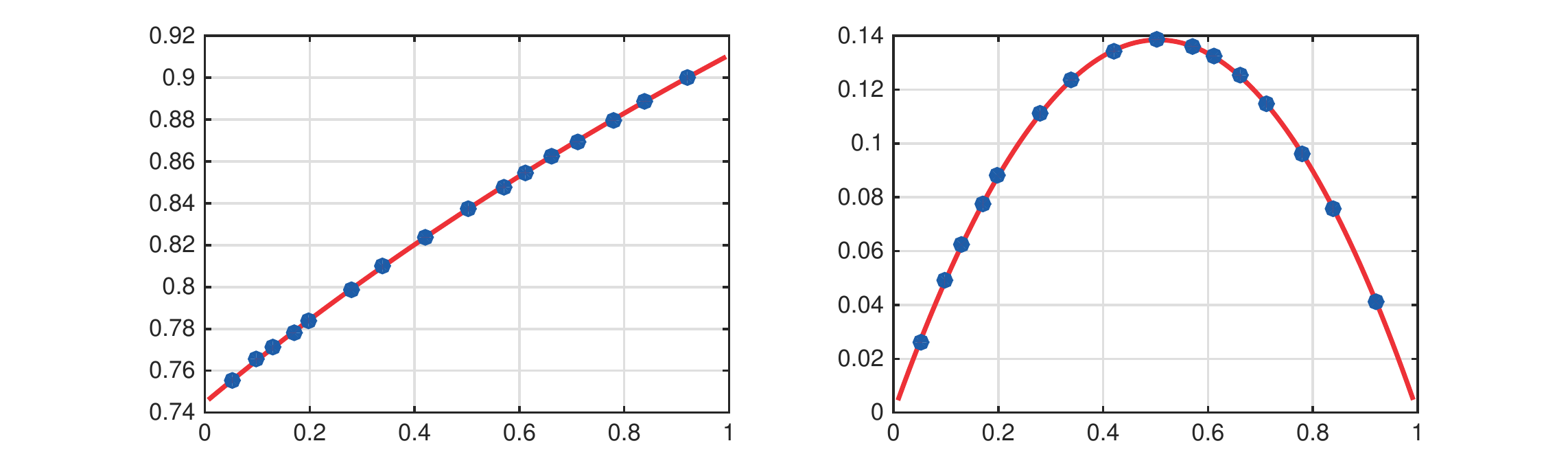}
\caption{Dynamic friction coefficient $A_{lh}$ (left) and thermo-diffusion ratio $k_T^{(i)}$ (right)  for the DT mixture. Solid red lines show the results obtained in Ref.~\cite{kagan2014thermo} with Zhdanov's formalism and blue circles show the corresponding results  obtained by digitizing Fig.~1 of  Ref.~\cite{simakov2016hydrodynamic-2}.}
\label{fig: diff} 
\end{figure}

Simakov and Molvig use different representation for the diffusive flux. In particular, they operate with the gradient of the number fraction of the lighter species $\nabla x$ instead of the mass fraction $\nabla c$. To set the correspondence between the two expressions for the diffusive flux we notice that 
\begin{equation}
\label{eq: grad-x}
\nabla x  = \frac{1}{m_l m_h} \Bigl(\frac{\rho}{n_i} \Bigr)^2 \nabla c,
\end{equation}
where $\rho$ and $n_i$ are the total mass and number densities of the ionic mixture, respectively. Then, it is straightforward to see that Simakov \& Molvig results for $k_p$, $k_E$ and $k_T^{(e)}$ are identical to Eqs.~(\ref{eq: baro-diff-ratio})-(\ref{eq: thermo-diff-ratio-el}). To compare $A_{lh}$ and $k_T^{(i)}$ we consider the DT case, for which Simakov \& Molvig results can be retrieved by digitizing the data from Fig.~1 of Ref.~\cite{simakov2016hydrodynamic-2}. We then plot them in Fig.~\ref{fig: diff} of this note over the corresponding results of Ref.~\cite{kagan2014thermo} to see that the predictions of the Simakov \& Molvig are again identical to the earlier results.



\providecommand{\newblock}{}

\end{document}